\documentclass[runningheads, envcountsame, a4paper]{llncs}

\usepackage{amsmath,amsfonts}
\usepackage[dvipsnames]{xcolor}
\usepackage[ruled,vlined,linesnumbered,longend]{algorithm2e}

\SetCommentSty{mycommfont}
\usepackage{graphicx}
\usepackage{caption}
\usepackage{subcaption}

\usepackage{comment}

\begin{document}

\title{Distributed Path Reconfiguration and Data Forwarding in Industrial IoT Networks}
\titlerunning{Path Reconfiguration and Data Forwarding in Industrial Networks}  % abbreviated title (for running head)

\author{Theofanis~P.~Raptis%\orcidID{0000-0002-2906-584X} 
\and
Andrea~Passarella \and
Marco~Conti 
}
\authorrunning{T.~P.~Raptis, A.~Passarella, and M.~Conti} % abbreviated author list (for running head)
%%%% list of authors for the TOC (use if author list has to be modified)
%\tocauthor{Ivar Ekeland, Roger Temam, Jeffrey Dean, David Grove,Craig Chambers, Kim B. Bruce, and Elisa Bertino}
\institute{National Research Council, Institute of Informatics and Telematics, Pisa, Italy\\
\email{\{theofanis.raptis, andrea.passarella, marco.conti\}@iit.cnr.it}
%,\\ WWW home page:\texttt{http://users/\homedir iekeland/web/welcome.html}
}

\maketitle              % typeset the title of the contribution

\begin{abstract}
In today's typical industrial environments, the computation of the data distribution schedules is highly centralised. Typically, a central entity configures the data forwarding paths so as to guarantee low delivery delays between data producers and consumers. However, these requirements might become impossible to meet later on, due to link or node failures, or excessive degradation of their performance. In this paper, we focus on maintaining the network functionality required by the applications after such events. We avoid continuously recomputing the configuration centrally, by designing an energy efficient local and distributed path reconfiguration method. Specifically, given the operational parameters required by the applications, we provide several algorithmic functions which locally reconfigure the data distribution paths, when a communication link or a network node fails. We compare our method through simulations to other state of the art methods and we demonstrate performance gains in terms of energy consumption and data delivery success rate as well as some emerging key insights which can lead to further performance gains.

\end{abstract}

\section{Introduction}

With the introduction of Internet of Things (IoT) concepts in industrial application scenarios, industrial automation is undergoing a tremendous change. This is made possible in part by recent advances in technology that allow interconnection on a wider and more fine-grained scale \cite{7883994}. The core of distributed automation systems and networks is essentially the reliable exchange of data. Any attempt to steer processes independently of continuous human interaction requires, in a very wide sense, the flow of data between some kind of sensors, controllers, and
actuators \cite{6042563}.

In today's typical industrial configurations, the computation of the data exchange and distribution schedules is quite primitive and highly centralised. Usually, the generated data are transferred to a central network controller or intermediate network proxy nodes using wireless links. The controller analyses the received information and, if needed, reconfigures the network paths, the data forwarding mechanisms, the caching proxy locations and changes the behaviour of the physical environment through actuator devices. Traditional data distribution schemes can be implemented over relevant industrial protocols and standards, like IEC WirelessHART, IEEE 802.15.4e and IETF 6TiSCH.

%IEC WirelessHART \cite{4550808}, IEEE 802.15.4e \cite{DEGUGLIELMO20161} and IETF 6TiSCH \cite{6979984}.

Those entirely centralised and offline computations regarding data distribution scheduling, can become inefficient in terms of energy, when applied in industrial IoT networks. In industrial environments, the topology and connectivity of the network may vary due to link and sensor-node failures \cite{7058728}. %Furthermore, sensors may also be subject to RF interference, highly caustic or corrosive environments, high humidity levels, vibrations, dirt and dust, or other conditions that challenge performance \cite{us2002}. 
Also, very dynamic conditions, which make communication performance much different from when the central schedule was computed, possibly causing sub-optimal performance, may result in not guaranteeing application requirements. These %harsh environmental conditions and 
dynamic network topologies may cause a portion of industrial sensor nodes to malfunction. With the increasing number of involved battery-powered devices, industrial IoT networks may consume substantial amounts of energy; more than needed if local, distributed computations were used. 

\textbf{Our contribution.} In this paper we consider an industrial IoT network comprised of sensor and actuator nodes. Data consumers (actuators) and producers (sensors) are known. A number of intermediate resource-rich nodes act as proxies. We assume that applications require a certain maximum delivery delay from proxies to consumers, and that, at some point in time, a central controller computes an optimal set of multi-hop paths from producers to proxies, and proxies to consumers, which guarantee a maximum delivery delay, while maximising the energy lifetime of the networks (i.e., the time until the first node in the network exhaust energy resources). We focus on maintaining the network configuration in a way such that application requirements are met after important network operational parameters change due to some unplanned events (e.g., heavy interference, excessive energy consumption), while guaranteeing an appropriate use of nodes energy resources. %In order to achieve this, we derive an energy efficient local and distributed path reconfiguration method. Specifically, 
We provide several efficient algorithmic functions which reconfigure the paths of the data distribution process, when a communication link or a network node fails. %, or its conditions deteriorate too much with respect to the point in time when the original configuration was centrally determined. 
The functions regulate how the local path reconfiguration should be implemented and how a node can join a new path or modify an already existing path, ensuring that there will be no loops. The proposed method can be implemented on top of existing data forwarding schemes designed for industrial IoT networks. We demonstrate through simulations the performance gains of our method in terms of energy consumption and data delivery success rate.% compared to other state of the art solutions.

\section{Related works}

%The design of industrial IoT architectures that take into account networking metrics such as energy and latency has recently been attracting much attention from both academia and industry \cite{7785890}, \cite{8311660}. However, most of those works describe in a high level the requirements for the operation of industrial networked systems. 

The previous works that are the most related to this paper are \cite{7876128}, \cite{8115846}, \cite{8241428} and \cite{draft}. Although those works proxy nodes for the efficient distributed management of network data, they all perform path selection computations centrally. Placement and selection strategies of caching proxies in industrial IoT networks have been investigated in \cite{7876128}. Efficient proxy-consumer assignments  are presented in \cite{8115846}. Data re-allocation methods among proxies for better traffic balancing are presented in \cite{8241428}. Scheduling of the data distribution process maximising the time until the first network node dies is suggested in \cite{draft}, respecting end-to-end data access latency constraints of the order of $100$ ms, as imposed by the industrial operators \cite{reqs}. Different to those works, in this paper, we present a method which exploits the local knowledge of the network nodes so as to perform distributed path reconfiguration computations towards more efficient energy dissipation across the network.

\section{The model}

We model an industrial IoT network as a graph $G=(V,E)$. Typically, the network features three types of devices \cite{6714496}: resource constrained sensor and actuator nodes $u \in V$, a central network controller $C$, and a set of proxy nodes in a set $P$, with $P \subset V$, $|P| \ll |V-P|$. Every node $u \in V$, at time $t$, has an available amount of finite energy $E^t_u$. In general, normal nodes $u$ have limited amounts of initial energy supplies $E^0_u$, and proxy nodes have significantly higher amounts of initial energy supplies, with $E^0_p \gg E^0_u, \forall u \in V, p \in P$. 

%\subsection{The network}

A node $u \in V$ can achieve one-hop data propagation using suitable industrial wireless technologies (e.g., IEEE 802.15.4e) to a set of nodes which lie in its neighbourhood $N_u$. $N_u$ contains the nodes $v \in V$ for which it holds that $\rho_u \geq \delta(u,v)$, where $\rho_u$ is the transmission range of node $u$ (defined by the output power of the antenna) and $\delta(u,v)$ is the Euclidean distance between $u$ and $v$. The sets $N_u$ are thus defining the set of edges $E$ of the graph $G$. Each one-hop data propagation from $u$ to $v$ results in a latency $l_{uv}$.  Assuming that all network nodes operate with the same output power, each one-hop data propagation from $u$ to $v$ requires and amount of $\epsilon_{uv}$ of energy dissipated by $u$ so as to transmit one data piece to $v$. A node can also transmit control messages to the network controller $C$ by consuming $\epsilon_{cc}$ amount of energy. For this kind of transmissions, we assume that more expensive wireless technology is needed, and thus we have that $\epsilon_{cc} \gg \epsilon_{uv}$ (for example, the former can occur over WiFi or LTE links, while the latter over 802.15.4 links). 

%\subsection{The data}

Occasionally, data generation occurs in the network, relevant to the industrial process. The data are modelled as a set of data pieces $D = \{D_i\}$. Each data piece is defined as $D_i = (s_i, c_i, r_i)$, where $s_i \in V$ is the source of data piece $D_i$, $c_i \in V$ is the consumer\footnote{If the same data of a source, e.g., $s_1$, is requested by more than one consumers, e.g., $c_1$ and $c_2$, we have two distinct data pieces, $D_1 = (s_1, c_1, r_1)$ and $D_2 = (s_2, c_2, r_2)$, where $s_1 = s_2$.} of data piece $D_i$, and $r_i$ is the data generation rate of $D_i$. Each data piece $D_i$ is circulated in the network through a multi-hop path $\Pi_{s_ic_i}$. Each node $u \in \Pi_{s_ic_i}$ knows which is the previous node $previous(i,u) \in \Pi_{s_ic_i}$ and the next node $next(i,u) \in \Pi_{s_ic_i}$ in the path of data piece $D_i$. Without loss of generality, we divide time in time cycles $\tau$ and we assume that the data may be generated (according to rate $r_i$) at each source $s_i$ at the beginning of each $\tau$, and circulated during $\tau$. The data generation and request patterns are not necessarily synchronous, and therefore, the data need to be cached temporarily for future requests by consumers. This asynchronous data distribution is usually implemented through an industrial pub/sub system \cite{8115846}. A critical aspect in the industrial operation is the timely data access by the consumers upon request, and, typically, the data distribution system must guarantee that a given maximum data access latency constraint (defined by the specific industrial process) is satisfied. We denote this threshold as $L_{\textnormal{max}}$.

%\subsection{The latency constraint}

Due to the fact that the set $P$ of proxy nodes is strong in terms of computation, storage and energy supplies, nodes $p \in P$ can act as proxy in the network and cache data originated from the sources, for access from the consumers when needed. This relieves the IoT devices from the burden of storing data they generate (which might require excessive local storage), and helps meeting the latency constraint. Proxy selection placement strategies have been studied in recent literature \cite{8115846,7876128}. We denote as $L_{uv}$ the latency of the multi-hop data propagation of the path $\Pi_{uv}$, where $L_{uv} = l_{ui} + l_{ii+1} + ... + l_{i+nv}$. %When a data piece $D_i$ is generated at source $s_i$, it is delivered and stored to a proxy $p$ via a multi-hop wireless path. 
Upon a request from $c_i$, data piece $D_i$ can be delivered from $p$ via a (distinct) multi-hop path. We denote as $L_{c_i}$ the data access latency of $c_i$, with $L_{c_i} = L_{c_ip} + L_{pc_i}$. We assume an existing mechanism of initial centrally computed configuration of the data forwarding paths in the network, e.g., as presented in \cite{draft}. In order to meet the industrial requirements the following constraint must be met: $L_{c_i} \leq L_{\textnormal{max}}, \forall c_i \in V$.

\section{Network epochs and their maximum duration}
%\subsection{The network epochs}

%In order to formulate the temporal aspects of the model we consider, we introduce the concept of epochs and we model the network as a time varying network \cite{2000klein,1337732}. 
In order to better formulate the data forwarding process through a lifetime-based metric, we define the network epoch. A network epoch $j$ is characterised by the time $J$ ($\tau$ divides $J$) elapsed between two consecutive, significant changes in the main network operational parameters. A characteristic example of such change is a sharp increase of $\epsilon_{uv}$ between two consecutive time cycles, due to sudden, increased interference on node $u$, which in turn leads to increased retransmissions on edge $(u,v)$ and thus higher energy consumption. In other words, $\frac{\epsilon_{uv}(\tau) - \epsilon_{uv}(\tau-1)}{\epsilon_{uv}(\tau)} > \gamma$, where $\gamma$ is a predefined threshold. During a network epoch, (all or some of) the nodes initially take part in a configuration phase (central or distributed), during which they acquire the plan for the data distribution process by spending an amount of $e_u^{\textnormal{cfg}}$ energy for communication. Then, they run the data distribution process. A network epoch is thus comprised of two phases: \emph{Configuration phase.} During this initial phase, the nodes acquire the set of neighbours from/to which they must receive/forward data pieces in the next epoch. \emph{Data forwarding phase.} During this phase the data pieces are circulated in the network according to the underlying network directives.

Network epochs are just an abstraction that is useful for the design and presentation of the algorithmic functions, but does not need global synchronisation. As it will be clear later on, each node locally identifies the condition for which an epoch is finished from its perspective, and acts accordingly. Different nodes ``see'' in general different epochs. Although some events which affect the epoch duration cannot be predicted and thus controlled, we are interested in the events which could be affected by the data distribution process and which could potentially influence the maximum epoch duration. We observe that an epoch cannot last longer than the time that the next node in the network dies. Consequently, if we manage to maximise the time until a node dies due to energy consumption, we also make a step forward for the maximisation of the epoch duration.

We now define the maximum epoch duration, as it can serve as a useful metric for the decision making process of the distributed path reconfiguration. The maximum epoch duration is the time interval between two consecutive node deaths in the network. Specifically, each epoch's duration is bounded by the lifetime of the node with the shortest lifetime in the network, given a specific data forwarding configuration. Without loss of generality, we assume that the duration of the configuration phase equals $\tau$. We define the variables, $x^{ij}_{uv}$ which hold the necessary information regarding the transmission of the data pieces across the edges of the graph. More specifically, for epoch $j$, $x^{ij}_{uv} = 1$ when edge $(u,v)$ is activated for data piece $D_i$. On the contrary, $x^{ij}_{uv}=0$ when edge $(u,v)$ is inactive for the transmission of data piece $D_i$. We denote as $a^j_{uv} = \sum_{i = 1}^{d} r_i x^{ij}_{uv}$ the aggregate data rate of $(u,v)$ for epoch $j$. Stacking all $a^j_{uv}$ together, we get $\mathbf{x}_u^j = [a^j_{uv}]$, the data rate vector of node $u$ for every $v \in N_u$. Following this formulation (and if we assumed that $J \rightarrow \infty$) the maximum lifetime of $u$ during epoch $j$ can be defined as:
%\begin{equation}
%T_u(\mathbf{x}_j) = \frac{E_u}{\sum_{v \in N_u} \epsilon_{uv} a^j_{uv}}.
%\label{eq::nodelife}
%\end{equation}

\begin{equation} 
T_u(\mathbf{x}_u^j) =
  \begin{cases}
    \frac{E^j_u}{\sum_{v \in N_u} \epsilon_{uv} a^j_{uv}}       & \quad \text{if } E^j_u > e_u^{\textnormal{cfg}}\\
    \tau  & \quad \text{if } E^j_u \leq e_u^{\text{cfg}}\\
    0  & \quad \text{if } E^j_u = 0
  \end{cases},
  \label{eq::eq}
\end{equation}
where $e_u^{\textnormal{cfg}}$ is the amount of energy that is needed by $u$ in order to complete the configuration phase. Consequently, given an epoch $j$, the maximum epoch duration is $J_{\text{max}} = \min_{u \in V} \left\{ T_u(\mathbf{x}_u^j) \mid \sum_{v \in N_u} x^{ij}_{uv} > 0\right\}$. 

There have already been works in the literature which identify, for each data source $s_i$, the proxy $p$ where its data should be cached, in order to maximise the total lifetime of the network until the first node dies \cite{draft} (or, in other words, maximise the duration of the first epoch: $\max \min_{u \in V} \left\{ T_u(\mathbf{x}_u^1) \mid \sum_{v \in N_u} x^{i1}_{uv} > 0\right\}$), and configure the data forwarding paths accordingly. Reconfigurations can be triggered also when the conditions under which a configuration has been computed, change. Therefore, (i) epoch duration can be shorter than $J$, and (ii) we do not need any centralised synchronisation in order to define the actual epoch duration. We consider the epoch as only an abstraction (but not a working parameter for the functions), which is defined as the time between two consecutive reconfigurations of the network,  following the functions presented in Section \ref{sec::pathrecon}.

%However, due to the fact that industrial IoT networks have to operate beyond this time point, regardless of some nodes failures, the initial configuration necessitates online reconfigurations on the data forwarding paths when different nodes die. %The aim of this paper is to ensure that we are able to %$\max \min_{u \in V} \left\{ T_u(\mathbf{x}_j) \mid \sum_{v \in N_u} x^{ij}_{uv} > 0\right\}$, $\forall j$, 
% efficiently handle the network reconfigurations.

\section{Path reconfiguration and data forwarding} \label{sec::pathrecon}

%When a significant change in the network happens due to interference, malfunction, or low energy supplies, some of the data distribution paths may get disconnected, or some nodes might start consuming much higher energy than expected. In this section we present the distributed path reconfiguration and data forwarding method. 
The main idea behind our method is the following: the nodes are initially provided with a centralised data forwarding plan. When a significant change in the network occurs, the nodes involved are locally adjusting the paths, using lightweight communication links among them (e.g., 802.15.4) instead of communicating with the central network controller (e.g., LTE, WiFi). The main metric used for the path adjustment is the epoch-related $T_u(\mathbf{x}_u^j)$, as defined in Eq.~\ref{eq::eq}.
\begin{comment}
The main functions that the method is using are summarised here, and then analysed more in-depth in the following subsections:

\begin{itemize}
\item \texttt{DistrDataFwd}$(u)$: The main function that each $u \in V$ is flashed with. It regulates the data forwarding process and calls the necessary path reconfiguration functions when needed.
\item \texttt{DataForwarding}$(\tau)$: Data forwarding schedule (e.g., \cite{draft}) computed offline and distributed to each $u \in V$ by $C$, at the first time cycle. Each $u$ forwards data during each $\tau$ according to the schedule. 
\item \texttt{Deactivate}$(u,v)$: Node $u$ deactivates the edge $(u,v)$ for every $D_i$.
\item \texttt{LocalPathConfig}$(u,v)$: Initiates a local path reconfiguration.
\item \texttt{JoinPath}$(i,w,v)$: Node $u$ establishes a path between nodes $w$ and $v$ for data piece $D_i$.
\item \texttt{ModifyPath}$(i,w,deleteArg,dirArg)$: Node $u$ modifies the path until $w$ joins the path, avoiding potential loops.
\item \texttt{local\_aodv+}$(u,v,TTL)$: Ad hoc on demand distance vector (AODV) path reconfiguration, ran locally with a short TTL route request messages, enhanced by taking into account also the estimated $T_u$ of the paths.
\item \texttt{Disconnect}($u$): Node $u$ terminates operation, deactivating $(u,v), \forall v \in N_u$.
\end{itemize}

\end{comment}
The functions' pseudocode is presented in the following subsections. Due to lack of space we omit the presentation of some functions' pseudocodes, but those can be found in the extended version of the paper \cite{extended}. The functions are presented in upright typewriter font and the messages which are being sent and received are presented in italics typewriter font. The arguments in the parentheses of the functions are the necessary information that the functions need in order to compute the desired output. The arguments in the brackets of the messages are the destination nodes of the messages and the arguments in the parentheses of the messages are the information carried by the messages. We assume that a node $u$ complies with the following rules: $u$ knows the positions of every $v \in N_u$, $u$ knows the neighbourhood $N_v$ of every node $v$ in its own neighbourhood $N_u$, and $u$ stores only local information or temporarily present data pieces in transit.

\begin{comment}
Messages:
\begin{itemize}
\item  \texttt{\emph{status}}[$C$]$(E_u, \epsilon_{uv})$
\item \texttt{\emph{plan}}[$u$]
\item \texttt{\emph{alert}}[$previous(i,u)$]$(u, v)$
\item \texttt{\emph{join}}[$u$]$(i,w,v)$
\item \texttt{\emph{modify\_path}}$[w](i,u,joinArg,deleteArg)$
\item  \texttt{\emph{explore}}[$w$]$(i,u,v,TTL)$
\item \texttt{\emph{explore\_report}}[$u$]$(i,w,list\_w)$
\end{itemize}

\end{comment}

\begin{algorithm}[t!]
\DontPrintSemicolon
%\SetKwInOut{Input}{Input}\SetKwInOut{Output}{Output}
%$\forall u \in V$ with $E_u > 0$: $u$ communicates with the central controller\;
\If{$E_u^0 >0$}{ \label{algo::DistrDataFwd::init1}
%\tcc{nodes communicate their status to $C$}
send \texttt{\emph{status}}[$C$]$(E_u, \epsilon_{uv}, l_{uv})$\;
%\tcc{$C$ computes paths and data storage locations (e.g., \cite{draft})}
receive \texttt{\emph{plan}}[$u$]\;%\tcc{$\tau=1,j=1$}
$\tau = 1$\; \label{algo::DistrDataFwd::init2}
%\tcc{$u$ repeats until out of energy or until extreme interference}
\Repeat{$E_u = 0$ or $\frac{\epsilon_{uv}(\tau) - \epsilon_{uv}(\tau-1)}{\epsilon_{uv}(\tau)} > \gamma$ for $>50\%$ of active edges $(u,v)$ of $u$}{ \label{algo::DistrDataFwd::rep1}
	run \texttt{DataForwarding}$(\tau)$\; \label{algo::DistrDataFwd::fwd}
%	\tcc{Deactivate single edge}
	\If{$\exists (u,v)$ with $\frac{\epsilon_{uv}(\tau) - \epsilon_{uv}(\tau-1)}{\epsilon_{uv}(\tau)} > \gamma$}{ \label{algo::DistrDataFwd::msgs1} \label{algo::DistrDataFwd::de1}
	\texttt{Deactivate}$(i,(u,v))$, $\forall D_i$\;
	send \texttt{\emph{alert}}[$previous(i,u)$]$(u, v)$, $\forall D_i$\;
	} \label{algo::DistrDataFwd::de2}
%	\tcc{Reconfigure path after alert reception}
	\If {receive \texttt{alert}[$u$]$(v, next(i, v))$}{ \label{algo::DistrDataFwd::lpc}
		\texttt{Deactivate}$(i,(u,v))$\;
		call \texttt{LocalPathConfig}$(i,u,next(i,v))$\;%\tcc{$\tau++,j++$}		
	}
%	\tcc{Join path}
	\If {receive \texttt{join}[$u$]$(i,w,v)$}{ \label{algo::DistrDataFwd::joi}
		call \texttt{JoinPath}$(i,w,v)$\;%\tcc{$\tau++,j++$}		
	}
%	\tcc{Modify existing path and avoid loops}
	\If{receive \texttt{modify\_path}$[u](i,w,deleteArg,dirArg)$}{ \label{algo::DistrDataFwd::mod}
		call \texttt{ModifyPath}$(i,w,deleteArg,dirArg)$\;
	} \label{algo::DistrDataFwd::msgs2}
$\tau++$\;
}\label{algo::DistrDataFwd::rep2}

}
%\tcc{Send alert and disconnect}
send \texttt{\emph{alert}}[$previous(i,u)$]$(u, v)$, $\forall D_i$, $\forall v \in N_u$\; \label{algo::DistrDataFwd::ciao}
\texttt{Disconnect}($u$)\;
\caption{\texttt{DistrDataFwd$(u)$}}
\label{algo::DistrDataFwd}
\end{algorithm}

\textbf{Distributed data forwarding.} The distributed data forwarding function \texttt{DistrDataFwd}$(u)$ pseudocode is being ran on every node $u$ of the network and is provided in the body of Alg.~\ref{algo::DistrDataFwd}. At first, if $E_u^0 > 0$, the node communicates its status to the central network controller (which uses the method presented in \cite{draft} for computing the data distribution parameters (proxy allocation, data forwarding paths) in an initial setup phase of the network), it receives the data forwarding plan and it initiates the first time cycle (lines \ref{algo::DistrDataFwd::init1}-\ref{algo::DistrDataFwd::init2}). Then, for every time cycle $u$ repeats the following process, until either it is almost dead, or more than half of its associated wireless links spend more energy compared to the previous time cycle, according to the system parameter $\gamma$ (lines \ref{algo::DistrDataFwd::rep1}-\ref{algo::DistrDataFwd::rep2}): $u$ starts the data forwarding process according to the data distribution plan received by $C$ (line \ref{algo::DistrDataFwd::fwd}). Afterwards, it checks if a set of control messages have been received from any $v \in N_u$ and acts accordingly, by calling the necessary functions (lines \ref{algo::DistrDataFwd::msgs1}-\ref{algo::DistrDataFwd::msgs2}). 

If $u$ detects that a link is consuming too much energy and has to be deactivated, it deactivates this link (by causing a path disconnection) and notifies the previous node in the path for $D_i$, $previous(i,u)$, by sending an alert message (lines \ref{algo::DistrDataFwd::de1}-\ref{algo::DistrDataFwd::de2}). For a given deactivated link $(u,v)$ for data piece $D_i$, alert messages contain information on which is the data piece of interest, and which were the two nodes $u,v$ in the path prior to disconnection. Then, $u$ checks whether there has been an alert message received (line \ref{algo::DistrDataFwd::lpc}), and calls function \texttt{LocalPathConfig} (displayed in Alg.~\ref{algo::LocalPathConfig}). Through this function the paths can be reconfigured accordingly, for all involved data pieces $D_i$. Due to the fact that \texttt{LocalPathConfig} sends some additional messages regarding joining a new path and modifying an existing one, $u$ then checks for reception of any of those messages (lines \ref{algo::DistrDataFwd::joi} and \ref{algo::DistrDataFwd::mod}) and calls the necessary functions \texttt{JoinPath} and \texttt{ModifyPath}.

Finally, $u$ sends an alert message to the previous nodes in the existing paths prior to final disconnection due to energy supplies shortage (line \ref{algo::DistrDataFwd::ciao}).

\textbf{Local path configuration.} A node $u$ calls the path configuration function \texttt{LocalPathConfig} when it receives an alert which signifies cease of operation of an edge $(u,v)$ due to a sudden significant increase of energy consumption due to interference $\left(\frac{\epsilon_{uv}(\tau) - \epsilon_{uv}(\tau-1)}{\epsilon_{uv}(\tau)} > \gamma \right)$ or a cease of operation of a node $v$ due to heavy interference in all of $v$'s edges or due to low energy supplies (Alg.~\ref{algo::DistrDataFwd}, lines \ref{algo::DistrDataFwd::de2} and \ref{algo::DistrDataFwd::ciao}). %\texttt{LocalPathConfig}$(u,v)$ is also called when $u$ detects that the link $(u,v)$ is draining too much energy. 

\begin{algorithm}[t!]
\DontPrintSemicolon
%\SetKwInOut{Input}{Input}\SetKwInOut{Output}{Output}
%\While {$ E_u > 0$ and $\epsilon_{uv} < \epsilon_{\text{max}}$}{

%	\For{$i=1:d$}{
%		run data distribution\;
%		\tcc{Node $v$ stopped operating: incr. interference/low batt.}
%		\If{$v \in N_u \cap \Pi_{s_i c_i}$ and $previous(i,v) = u$}{
%		\tcc{Replace $v$ with best lifetime $w$ and send join message to $w$} 
			\uIf{$\exists \iota \in N_u$ with $v \in N_{\iota}$ and $l_{u\iota} + l_{\iota v} \leq l_{uprevious(i,v)} + l_{previous(i,v)v}$}{ \label{algo::LocalPathConfig::chk}
				$w = \arg\max_{\iota \in N_u} T_{\iota}(\mathbf{x}_u^j)$\; \label{algo::LocalPathConfig::rpl}
				send \texttt{\emph{join}}[$w$]$(i,u,v)$\; \label{algo::LocalPathConfig::send}
				$\Pi_{s_i c_i} \leftarrow$ replace $v$ with $w$\;
				%\If{receive \texttt{\emph{modify\_path}}$[u](i,w,joinArg,deleteArg,dirArg)$}{
				%	\tcc{If no loop exists or forward loop exists}
				%	\uIf{$joinArg = joinYES$ and $deleteArg = deleteNO$}{
				% 		$\Pi_{s_i c_i} \leftarrow$ replace $v$ with $w$\;
				%	}
				%	\tcc{If backward loop exists}
				%	\ElseIf{$joinArg = joinYES$ and $deleteArg = deleteYES$}{
			%
				%	}
				%}
			}
%			\tcc{Explore extended neighbourhood and replace $v$ with $w,w'$}
			\Else{
				run \texttt{local\_aodv+}$(u,v,TTL)$ \label{algo::LocalPathConfig::aodv}
				
				}
%		}
		%\If {\texttt{\emph{received\_update}}$(v', next(i, v))$}{
		
		%}
%	}
%}
%\texttt{\emph{send\_alert}}$(u, next(i, u))$
\caption{\texttt{LocalPathConfig}$(i,u,v)$}
\label{algo::LocalPathConfig}
\end{algorithm}

 \texttt{LocalPathConfig} is inherently local and distributed. The goal of this function is to restore a functional path between nodes $u$ and $v$ by replacing the problematic node $previous(v)$ with a better performing node $w$, or if $w$ does not exist, with a new efficient multi-hop path $\Pi_{uv}$. At first, $u$ checks if there are nodes $\iota$ in its neighbourhood $N_u$ which can directly connect to $v$ and achieve a similar or better one-hop latency than the old configuration (line \ref{algo::LocalPathConfig::chk}). If there are, then the $w$ selected is the node $\iota$ which given the new data piece, will achieve a maximum lifetime compared to the rest of the possible replacements, i.e., $w = \arg\max_{\iota \in N_u} T_\iota(\mathbf{x}_u^j)$, and an acceptable latency $l_{uw} + l_{wv}$ (line \ref{algo::LocalPathConfig::rpl}). $u$ then sends to $w$ a \texttt{\emph{join}} message (line \ref{algo::LocalPathConfig::send}).
 
If such a node does not exist, then $u$ runs \texttt{local\_aodv+}, a modified, local version of AODV protocol for route discovery, between nodes $u$ and $v$. \texttt{local\_aodv+} is able to add more than one replacement nodes in the path. The main modification of \texttt{local\_aodv+} with respect to the traditional AODV protocol is that \texttt{local\_aodv+} selects the route which provides the  maximum lifetime $T_w(\mathbf{x}_u^j)$ for the nodes $w$ which are included in the route. Specifically, this modification with respect to the classic AODV is implemented as follows: The nodes piggyback in the route request messages the minimum lifetime $T_w(\mathbf{x}_u^j)$ that has been identified so far on the specific path. Then when the first route request message arrives at $v$, instead of setting this path as the new path, $v$ waits for a predefined timeout for more route request messages to arrive. Then, $v$ selects the path which provided the $\max\min_{w \in N_u} T_w(\mathbf{x}_u^j)$. The reader can find more details about the AODV protocol in \cite{aodv}.

\begin{figure}[t!]
\centering
    \begin{subfigure}[b]{0.32\textwidth}
    \centering
        \includegraphics[width=\columnwidth]{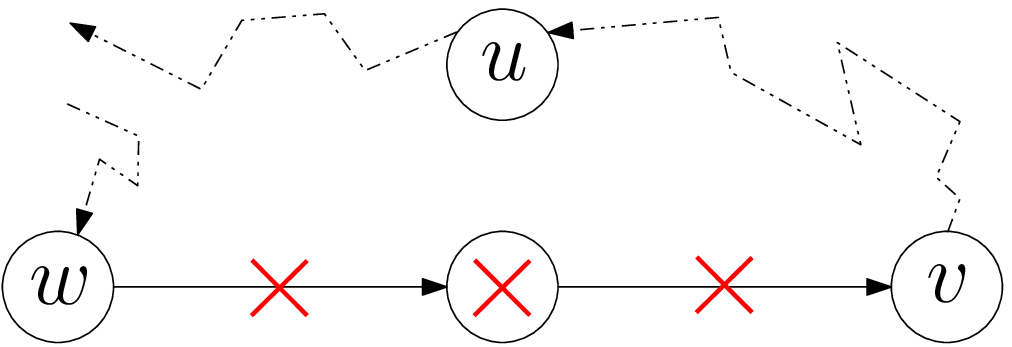}
        \caption{Unplanned change.}
        \label{fig::fwdloop1}
    \end{subfigure}
    \begin{subfigure}[b]{0.32\textwidth}
    \centering
        \includegraphics[width=\columnwidth]{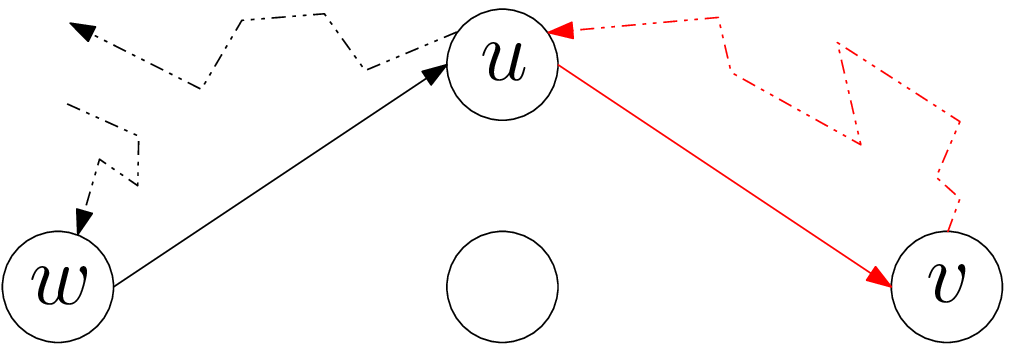}
        \caption{$u$ (re-)joining the path.}
        \label{fig::fwdloop2}
    \end{subfigure}   
    \begin{subfigure}[b]{0.32\textwidth}
    \centering
        \includegraphics[width=\columnwidth]{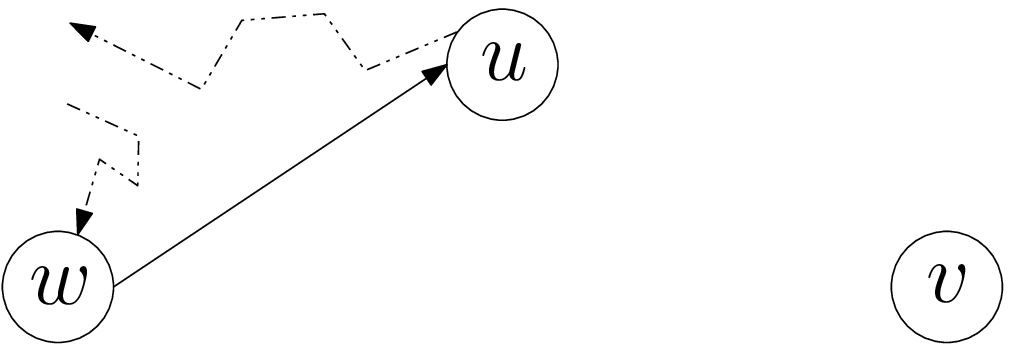}
        \caption{Loop elimination.}
        \label{fig::fwdloop3}
    \end{subfigure}   
\caption{Loop avoidance - forward loop.}
\label{fig::fwdloop}
\end{figure}

\begin{figure}[t!]
\centering
    \begin{subfigure}[b]{0.32\textwidth}
    \centering
        \includegraphics[width=\columnwidth]{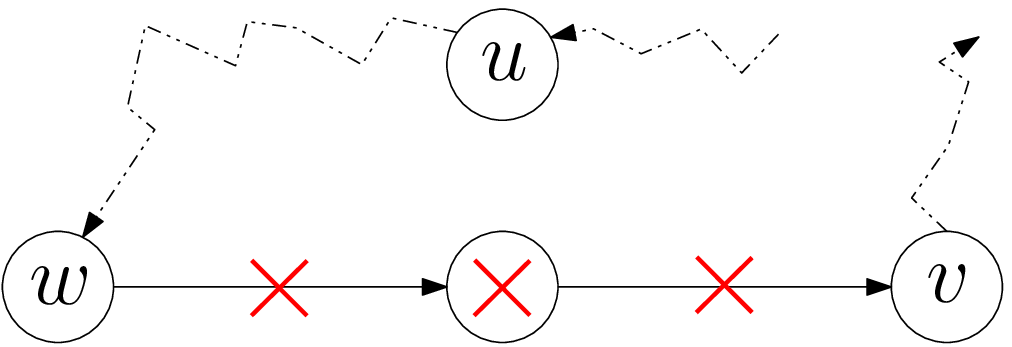}
        \caption{Unplanned change.}
        \label{fig::bwdloop1}
    \end{subfigure}
    \begin{subfigure}[b]{0.32\textwidth}
    \centering
        \includegraphics[width=\columnwidth]{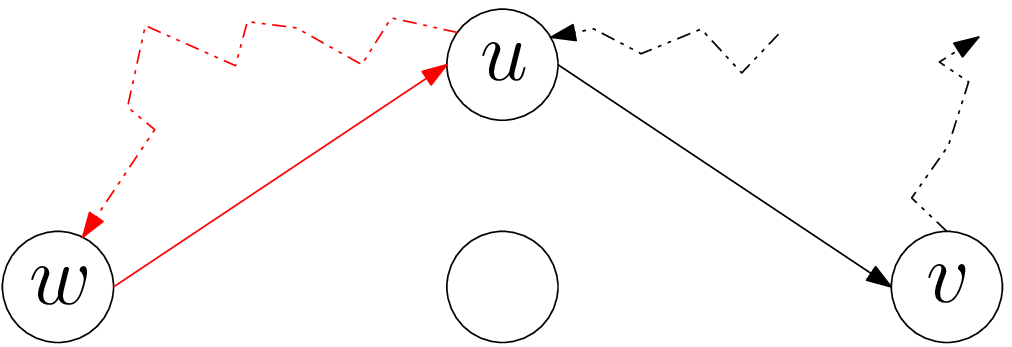}
        \caption{$u$ (re-)joining the path.}
        \label{fig::bwdloop2}
    \end{subfigure}   
    \begin{subfigure}[b]{0.32\textwidth}
    \centering
        \includegraphics[width=\columnwidth]{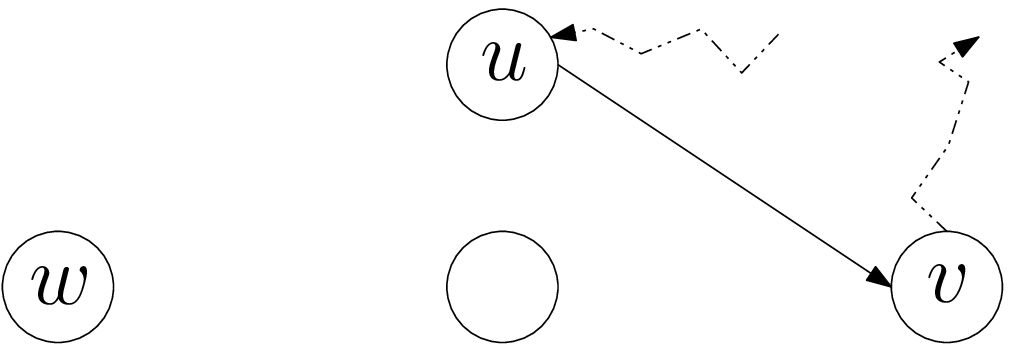}
        \caption{Loop elimination.}
        \label{fig::bwdloop3}
    \end{subfigure}   
\caption{Loop avoidance - backward loop.}
\label{fig::bwdloop}
%    \vspace{-0.7cm}
\end{figure}

\textbf{Joining new paths, modifying existing paths and avoiding loops.} In this subsection, we briefly describe the functions regarding joining a new path and modifying an already existing path for loop elimination. Due to lack of space, we do not include the pseudocode of those functions; however, they can be found at the extended version of this paper \cite{extended}. \texttt{JoinPath}$(i,w,v)$ is the function which regulates how, for data piece $D_i$, a node $u$ will join an existing path between nodes $w$ and $v$ and how $u$ will trigger a path modification and avoid potential loops which could result in unnecessary traffic in the network. Due to the fact that the reconfigurations do not use global knowledge, we can have three cases of $u$ joining a path: (i) $u$ is not already included in the path ($u \notin \Pi_{s_i c_i}$), (ii) $u$ is already included in the path ($u \in \Pi_{s_i c_i}$), and $w$ is preceding $u$ in the new path ($previous(i,u) = w$) with a new link $(w,u)$, and (iii) $u$ is already included in the path ($u \in \Pi_{s_i c_i}$), and $u$ is preceding $w$ in the new path ($previous(i,w) = u$) with a new link $(u,w)$. In all three cases, \texttt{JoinPath} sends a modification message to the next node to join the path, with the appropriate arguments concerning the deletion of parts of the paths, and the direction of the deletion, for avoidance of potential loops (see \cite{extended}). This messages triggers the function \texttt{ModifyPath} (see \cite{extended}). In case (i) it is apparent that there is no danger of loop creation, so there is no argument for deleting parts of the path. In order to better understand cases (ii) and (iii) we provide Figures \ref{fig::fwdloop} and \ref{fig::bwdloop}. In those Figures we can see how the function \texttt{ModifyPath} eliminates newly created loops on $u$ from path reconfigurations which follow unplanned network changes.

Following the loop elimination process, loop freedom is guaranteed for the cases where there are available nodes $w \in N_u$ which can directly replace $v$. In the case where this is not true and \texttt{LocalPathConfig} calls \texttt{local\_aodv+} instead (Alg.~\ref{algo::LocalPathConfig}, line \ref{algo::LocalPathConfig::aodv}), then the loop freedom is guaranteed by the AODV path configuration process, which has been proven to be loop free \cite{5371479}.

\section{Performance evaluation}

\begin{comment}

\begin{table}[t!]
\centering
\caption{Simulation parameters.}
\begin{tabular}{ l | l }\hline
\textbf{Parameter} & \textbf{Value}\\\hline
\multicolumn{2}{c}{\textbf{Topology}} \\\hline
deployment dimensions (2D grid) & $7.5$ m $\times$ $16.0$ m\\
$|V|$, $|E|$, $|P|$,  $\delta (u,v)$ & $18, 47, 4$, $2.5 - 2.8$\\
transmission range $\rho_u$, neighborhood $N_u$ & $3$ m, $v$, with $d(u,v) \leq 3$ m\\\hline
\multicolumn{2}{c}{\textbf{Time}} \\\hline
$\tau$, $L_{\text{max}}$, $TTL$ (\texttt{local\_aodv+}) & $1$ sec, $100$ ms, $2$ hops \\
experiment duration & $2000$ hours\\\hline
\multicolumn{2}{c}{\textbf{Data}}\\\hline
percentage of consumers $c_i$ & $0.05 - 45\%$\\
data piece generation rate $r_i$ & $1-8$ $D_i/\tau$\\
data piece size (incl. headers and CRC) & $9$ bytes\\\hline
\multicolumn{2}{c}{\textbf{Hardware assumptions}} \\\hline
MCU (e.g., MSP430), antenna (e.g., CC2420) & ultra low-power, IEEE 802.15.4  \\
max. battery capacity, initial energies $E_u^0$, $E_p^0$ & $830$ mAh / $3.7$ V, $0-1$ and $3$ Wh\\
transmission power for $e_{uv}$, for $e_{cc}$ & $-25$ dBm, $15$ dBm\\\hline
\end{tabular}
\label{tab::parameters}
\end{table}

\end{comment}

\begin{figure}[t!]
\centering
    \begin{subfigure}[b]{0.32\columnwidth}
    \centering
        \includegraphics[width=\columnwidth]{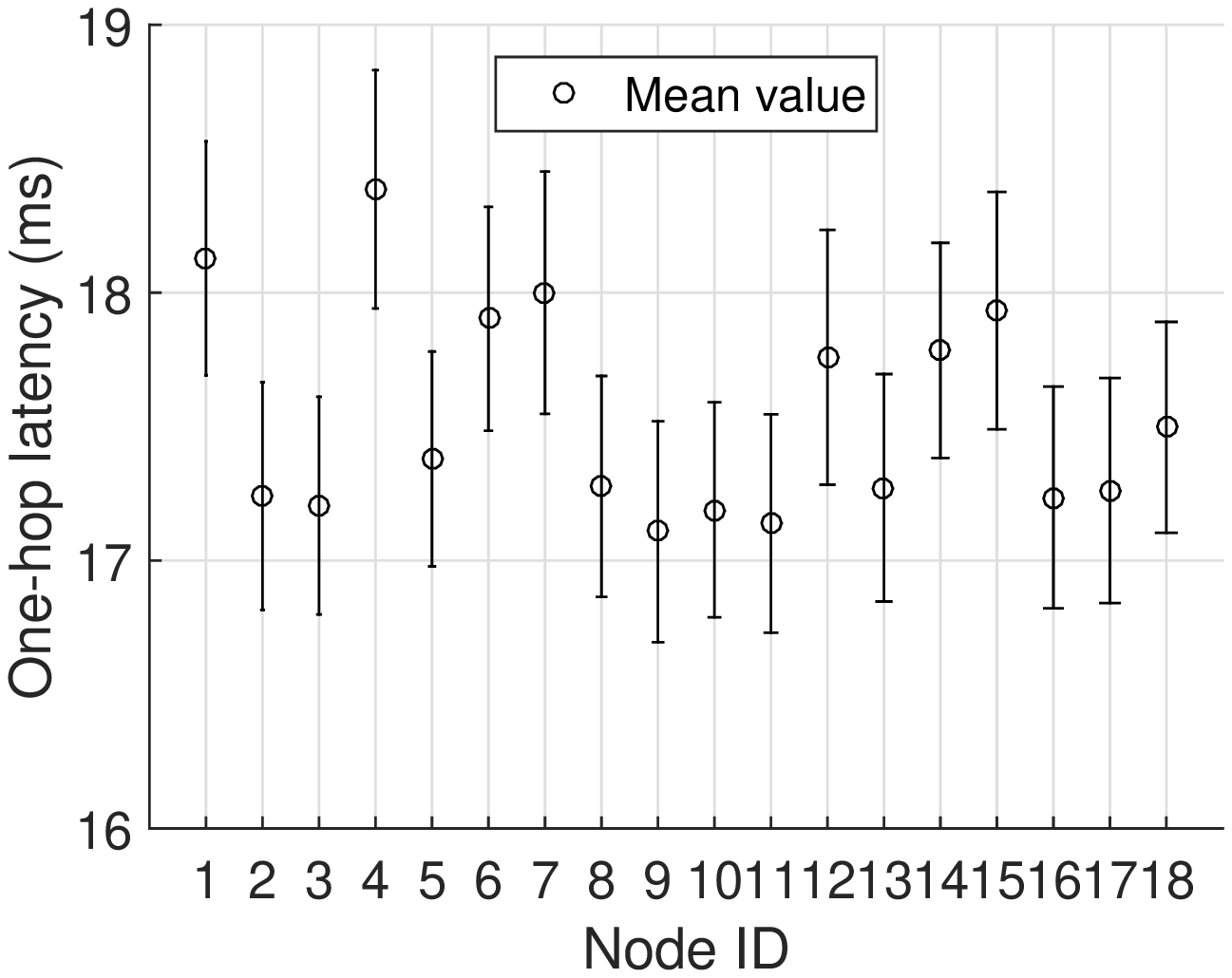}
        \caption{Device latencies.}
        \label{fig::error}
    \end{subfigure}
    \begin{subfigure}[b]{0.32\columnwidth}
    \centering
        \includegraphics[width=\columnwidth]{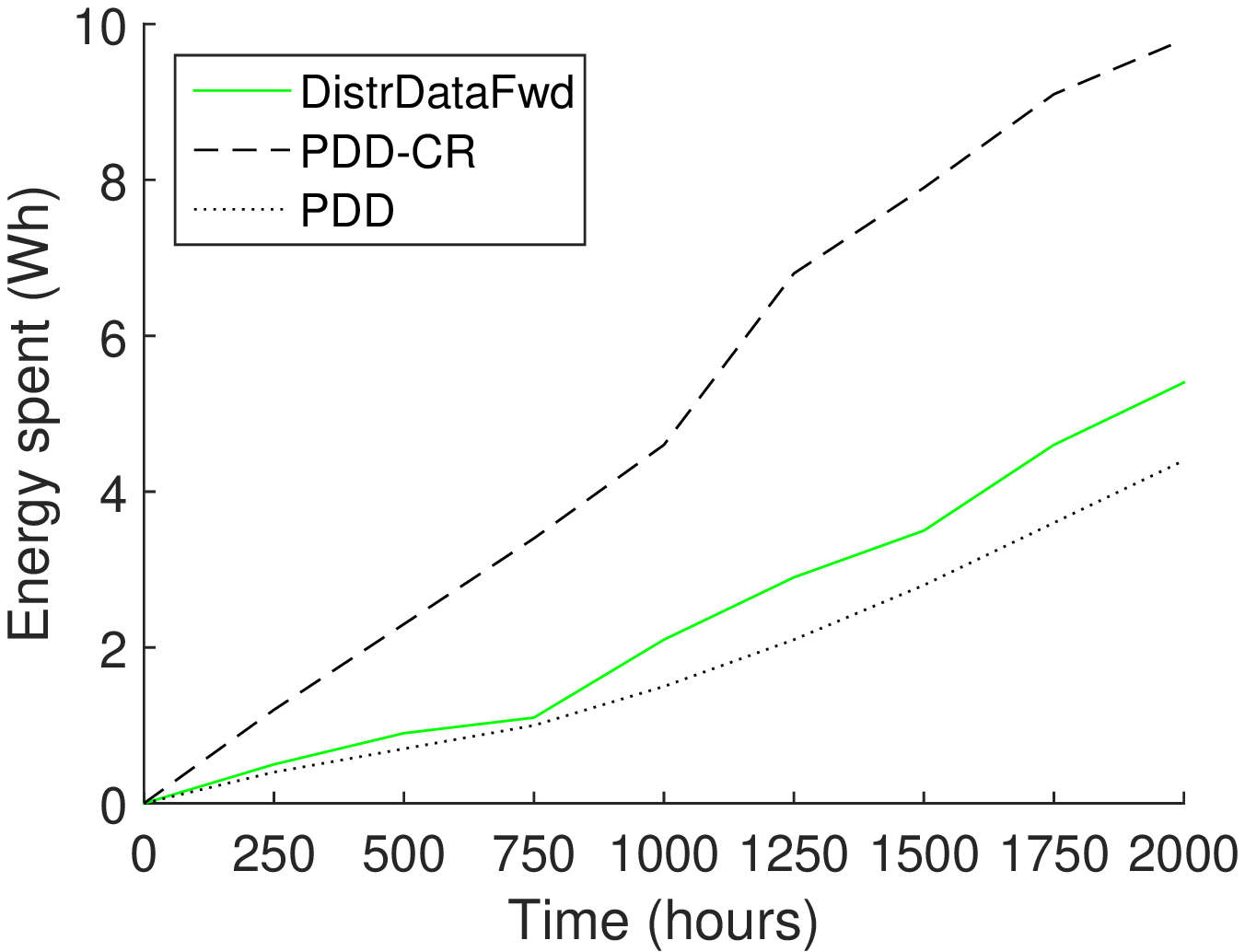}
        \caption{Energy consumption.}
        \label{fig::energy}
    \end{subfigure}
    \begin{subfigure}[b]{0.32\columnwidth}
    \centering
        \includegraphics[width=\columnwidth]{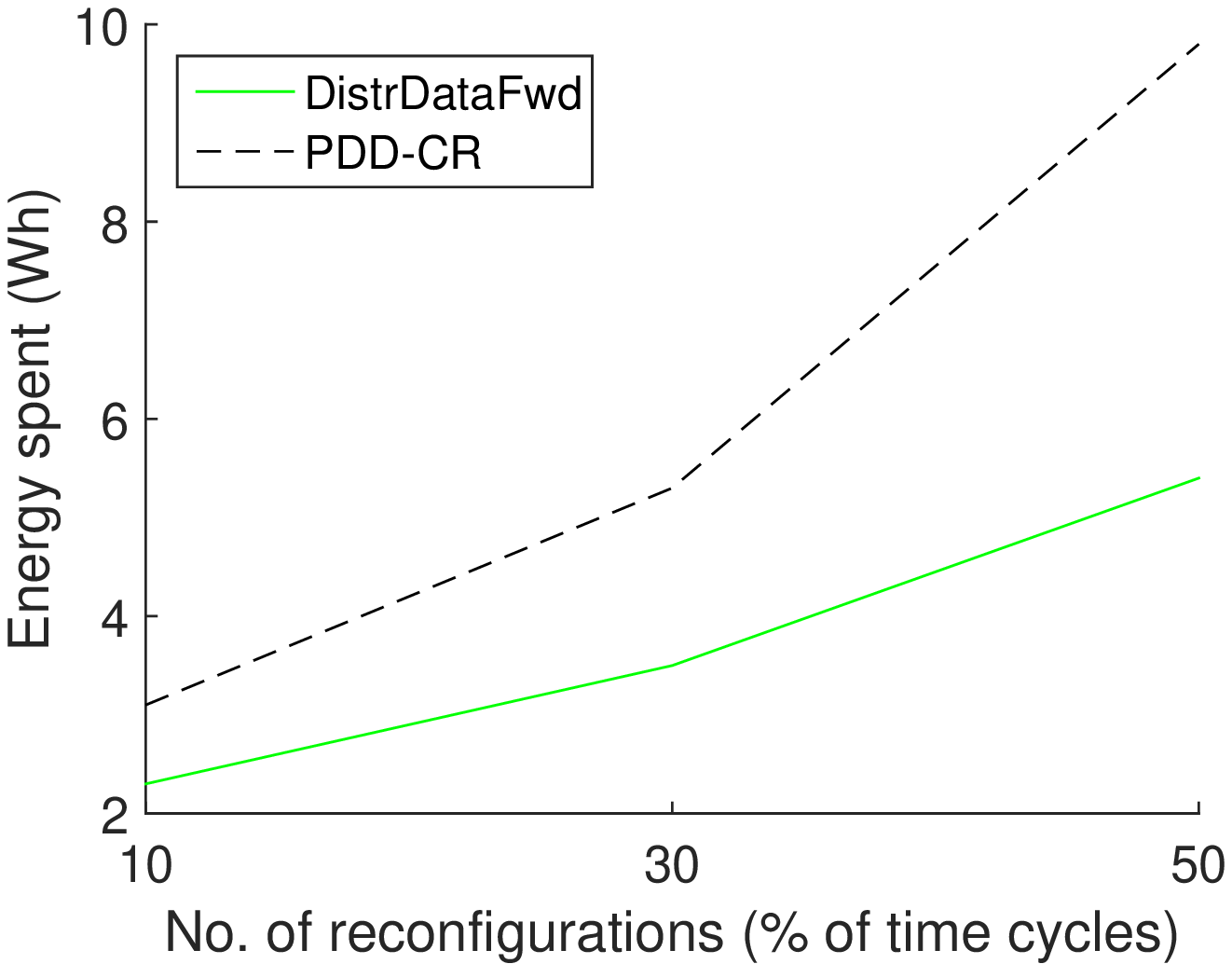}
        \caption{Var. reconfigurations}
        \label{fig::reconfig}
    \end{subfigure}

    \begin{subfigure}[b]{0.32\columnwidth}
        \centering
        \includegraphics[width=\columnwidth]{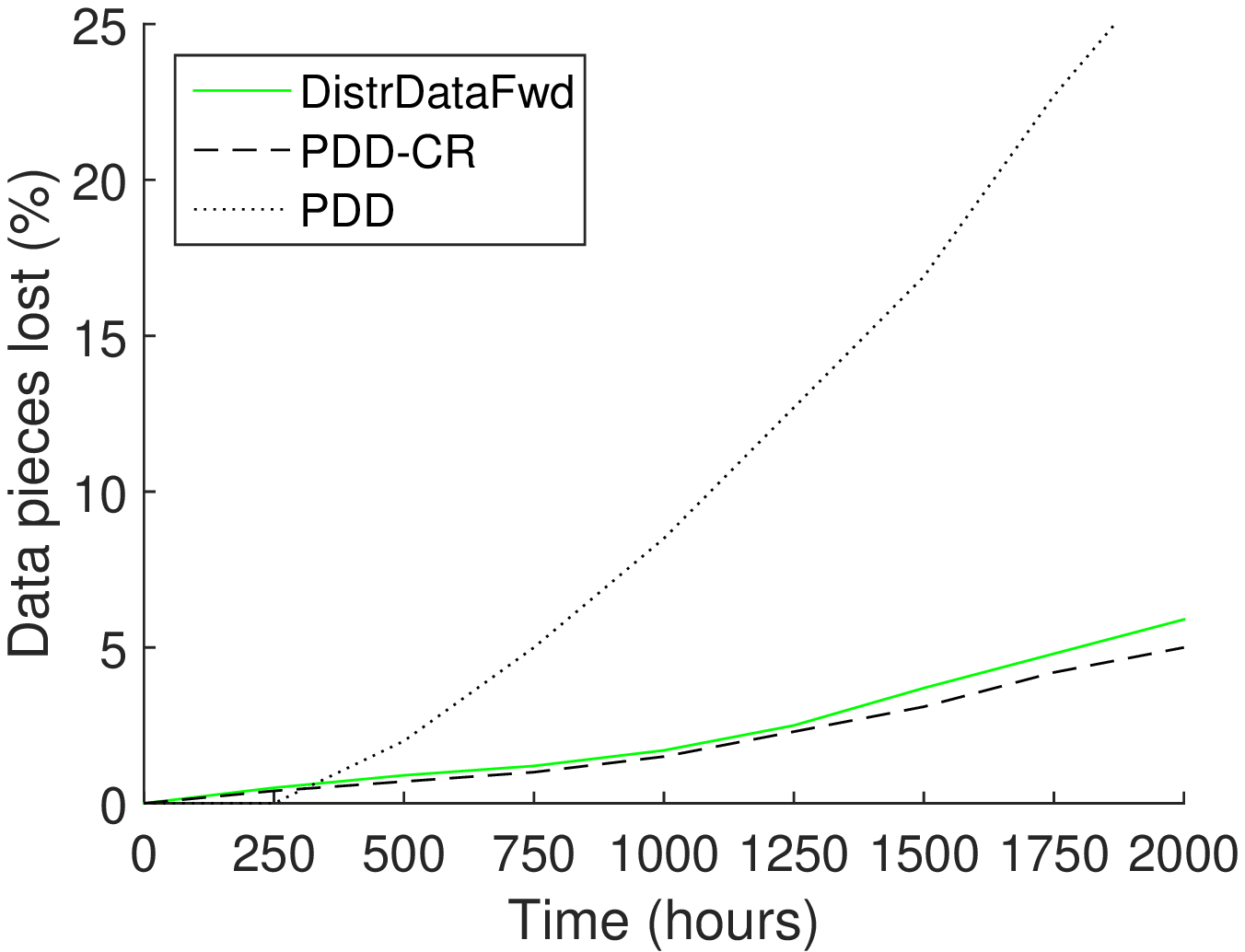}
        \caption{Data pieces lost.}
        \label{fig::data}
    \end{subfigure}
    \begin{subfigure}[b]{0.32\columnwidth}
    \centering
        \includegraphics[width=\columnwidth]{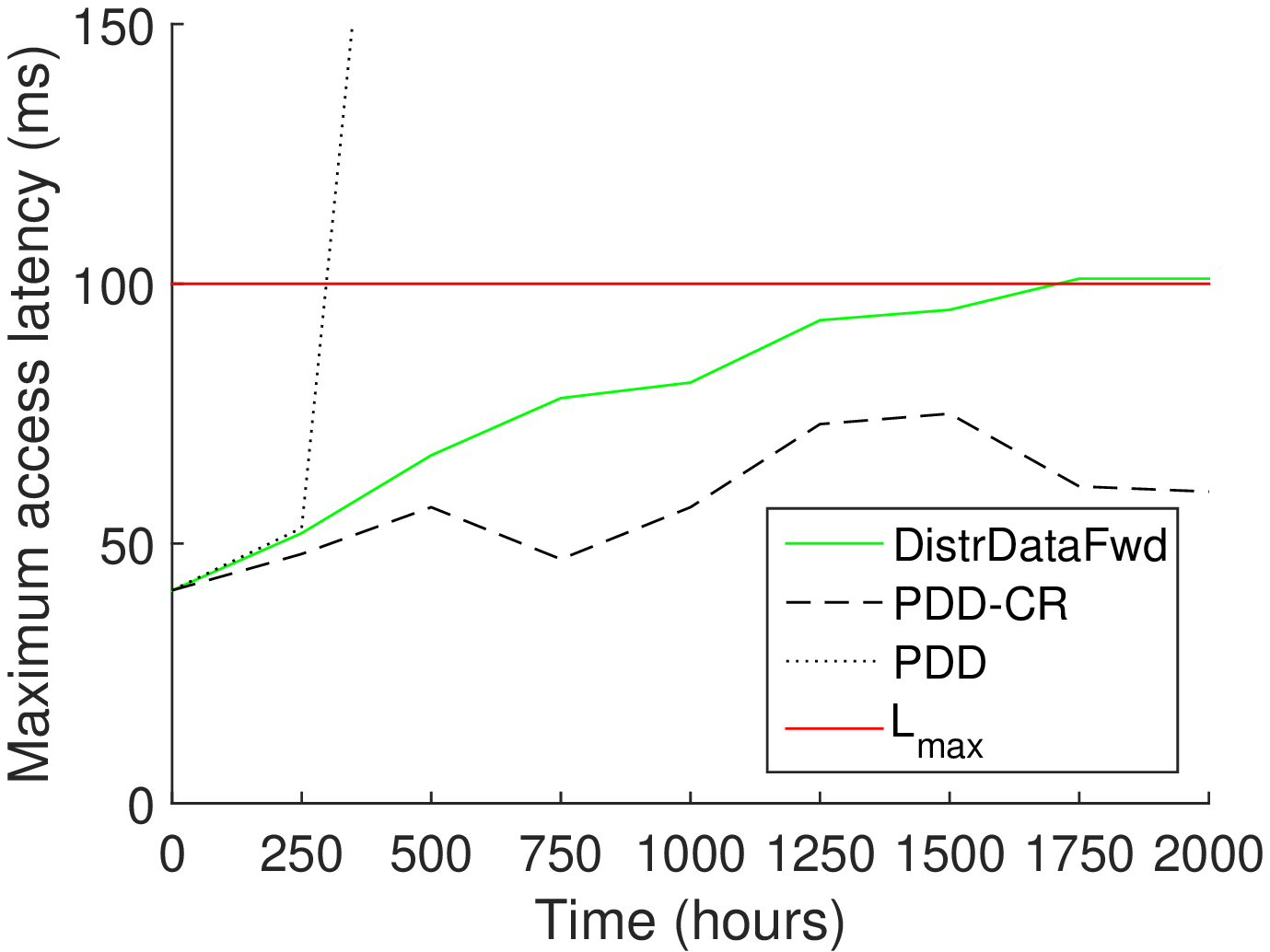}
        \caption{Max. access latency.}
        \label{fig::latency}
    \end{subfigure}
    \caption{Performance results.}\label{fig::results}
%    \vspace{-0.7cm}
\end{figure}

We implemented \texttt{DistrDataFwd} method and we conducted simulations in order to demonstrate its performance. We configured the simulation environment (Matlab) according to realistic parameters and assumptions. A table presenting the parameter configuration in details can be found in the extended version of the paper \cite{extended}. Briefly, we assume an industrial IoT network, comprised of devices equipped with ultra low-power MCUs like MSP430 and IEEE 802.15.4 antennae like CC2420, able to support industrial IoT standards and protocols like WirelessHART and IEEE 802.15.4e. We assume a structured topology (as in usual controlled industrial settings) of 18 nodes with 4 proxies which form a 2D grid with dimensions of $7.5$ m $\times$ $16.0$ m. We set the transmission power of the nodes for multi-hop communication to $-25$ dBm (typical low-power) which results in a transmission range of $3$ m. For the more expensive communication with the network controller, we set the transmission power to $15$ dBm, typical of wireless LAN settings. We set the time cycle $\tau = 1$ second, the percentage of consumers over the population $0.05-45$\% and we produce $1-8$ $D_i/\tau$ per consumer. In order to perform the simulations in the most realistic way, we align the $L_{\text{max}}$ value with the official requirements of future network-based communications for Industry 4.0 \cite{reqs}, and set the latency threshold to $L_{\text{max}} = 100$ ms. We set $\gamma = 50\%$, the $TTL$ argument of \texttt{local\_aodv+} equal to $2$, we assume a maximum battery capacity of $830$ mAh (3.7 V) and equip the nodes with energy supplies of $E_u^0 = 0-1$ Wh and $E_p^0 = 3$ Wh. Last but not least, in order to have a realistic basis for the one-hop latencies $l_{uv}$ to be used in the simulations, we performed some limited experiments with real devices. Specifically, we repeated one-hop propagation measurements multiple times for 18 WSN430 nodes, for different pairs of transmitting and receiving nodes. We concluded to the measurements that are shown in Fig.~\ref{fig::error}, after measuring the relevant latencies using WSN430 with CC2420 antenna, under the TinyOS operating system. We assigned to the different $l_{uv}$ the corresponding, variable intervals of values.

In order to have a benchmark for our method, we compared its performance to the performance of the \texttt{PDD} data forwarding method which was provided in \cite{draft}. Due to the fact that \texttt{PDD} was designed for static environments without significant network parameter changes, we also compare to a modified version of \texttt{PDD}, which incorporates central reconfiguration when needed (we denote this version as \texttt{PDD-CR}). Specifically, \texttt{PDD-CR} runs \texttt{PDD} until time $t$, when a significant change in the network happens, and then, all network nodes communicate their status ($E^t_u, e_{uv}, l_{uv}$) to the network controller $C$ by spending $e_{cc}$ amount of energy. $C$ computes centrally a new (near-optimal as shown in \cite{draft}) data forwarding plan and the nodes run the new plan. In our case, we run the \texttt{PDD-CR} reconfigurations for each case where we would do the same if we were running \texttt{DistrDataFwd}. As noted before, the conditions that trigger a change of the forwarding paths are either node related (a node dies) or link related (change of interference which results in $\frac{\epsilon_{uv}(\tau) - \epsilon_{uv}(\tau-1)}{\epsilon_{uv}(\tau)} > \gamma$)\footnote{The qualitative behaviour would not change in case of additional reconfiguration events, which simply increase the number of reconfigurations.}. We ran the experiments multiple times and we depict the average values so as to capture stochastic variations (e.g., different latencies on the same node, see Fig.~\ref{fig::error}).

\textbf{Energy efficiency.} The energy consumption over the entire network during 2000 hours of operation is depicted in Fig.~\ref{fig::energy}. The energy consumption values include the energy consumed for both the data distribution process and the reconfiguration. Our method achieves comparable energy consumption as \texttt{PDD}, despite being a local, adaptive method. This is explained by the following facts: \texttt{PDD-CR} requires more energy than \texttt{DistrDataFwd} for the path reconfiguration process, as during each epoch alteration every node has to spend $e_{cc}$ amount of energy for the configuration phase. On the contrary, in the \texttt{DistrDataFwd} case, only some of the nodes have to participate in a new configuration phase (usually the nodes in the neighbourhood of the problematic node), and spend significantly less amounts of energy. In the case of \texttt{PDD}, the nodes do not participate in configuration phases, so they save high amounts of energy. In Fig.~\ref{fig::reconfig}, we can also see the energy consumption of \texttt{DistrDataFwd} and \texttt{PDD-CR} for different percentages of reconfigurations (w.r.t. the number of time cycles $\tau$). It is clear that the more the reconfigurations that we have in the network, the more the gap between the performance of \texttt{DistrDataFwd} and \texttt{PDD-CR} increases.

\textbf{Data delivery rate.} The data pieces lost during 2000 hours of operation are depicted in Fig.~\ref{fig::data}. We consider a data piece as lost when the required nodes or path segments are not being available anymore so as to achieve a proper delivery. When a data piece is delivered, but misses the deadline $L_{\text{max}}$, it is not considered as lost, but we measure the high delivery latency instead. We can see that the low energy consumption of the \texttt{PDD} method comes at a high cost: it achieves a significantly lower data delivery rate than the \texttt{PDD-CR} and the \texttt{DistrDataFwd} methods. This is natural, because as noted before, \texttt{PDD} computes an initial centralised paths configuration and follows it throughout the entire data distribution process. The performance of the \texttt{DistrDataFwd} method stays very close to the performance of the \texttt{PDD-CR} method, which demonstrates the efficiency of \texttt{DistrDataFwd} in terms of successfully delivering the data pieces.

\textbf{Maximum data access latency.} The maximum data access latency during 2000 hours of operation is depicted in Fig.~\ref{fig::latency}. The measured value is the maximum value observed among the consumers, after asynchronous data requests to the corresponding proxies. \texttt{PDD} does not perform well, due to the fact that it is prone to early disconnections without reconfiguration functionality. The fluctuation of \texttt{PDD-CR}'s curve is explained by the re-computation from scratch of the data forwarding paths which might result in entirely new data distribution patterns in the network. \texttt{DistrDataFwd} respects the $L_{\text{max}}$ threshold for most of the time, however at around 1700 hours of network operation it slightly exceeds it for a single proxy-consumer pair. On the contrary, \texttt{PDD-CR} does not exceed the threshold. This performance is explained by the fact that \texttt{DistrDataFwd}, although efficient, does not provide any strict guarantee for respecting $L_{\text{max}}$, for all proxy-consumer pairs, mainly due to the absence of global knowledge on the network parameters during the local computations. \texttt{PDD-CR}, with the expense of additional energy for communication, is able to centrally compute near optimal paths and consequently achieve the desired latency. There are two simple ways of improving \texttt{DistrDataFwd}'s performance in terms of respecting the $L_{\text{max}}$ threshold: (i) insert strict latency checking mechanisms in the \texttt{local\_aodv+} function, with the risk of not finding appropriate (in terms of latency) path replacements, and thus lowering the data delivery ratio due to disconnected paths, and (ii) increase the $TTL$ argument of \texttt{local\_aodv+}, with the risk of circulating excessive amounts of route discovery messages, and thus increasing the energy consumption in the network. Including those mechanisms is left for future work.

\section{Conclusion}

We identified the need for a distributed reconfiguration method for data forwarding paths in industrial IoT networks. Given the operational parameters the network, we provided several efficient algorithmic functions which reconfigure the paths of the data distribution process, when a communication link or a network node fails. The functions regulate how the local path reconfiguration is implemented, ensuring that there will be no loops. We demonstrated the performance gains of our method in terms of energy consumption and data delivery success rate compared to other state of the art solutions.

\end{document}